\documentclass[prb,aps,showpacs,superscriptaddress,floatfix]{revtex4}
\usepackage{graphicx}
\usepackage{bm}
\usepackage{amsmath}
\begin{document}
%%%%%%%%%%%%%%%%%%%%
% Title
%%%%%%%%%%%%%%%%%%%%
\title{
Mechanism of $d_{x^2-y^2}$-wave superconductivity 
based on doped hole induced spin texture 
in high $T_c$ cuprates
}
\author{T. Morinari}
\affiliation{Yukawa Institute for Theoretical Physics, Kyoto
University, Kyoto 606-8502, Japan}
\date{\today}
\begin{abstract}
A mechanism of $d_{x^2-y^2}$-wave superconductivity is proposed for
the high-$T_c$ cuprates based on a spin texture with non-zero
topological density induced by doped holes through Zhang-Rice singlet
formation.
The pairing interaction arises from 
the magnetic Lorentz force like interaction between the holes and the
spin textures.
The stability of the pairing state against 
the vortex-vortex interaction and the Coulomb repulsion
is examined.
The mechanism suggests appearance of a p-wave pairing component
by introducing anisotropy in the CuO$_2$ plane.
\end{abstract}
%\pacs{ Valid PACS appear here.
%{\tt$\backslash$\string pacs\{\}} should always be input, even if
%empty.}

%\begin{multicols}{2}

%\narrowtext

\maketitle

\section{Introduction}
High-temperature superconductivity occurs in a moderately hole doped
Mott insulator.\cite{ANDERSON}
The symmetry of the Cooper pairs has been established to be
$d_{x^2-y^2}$-wave.\cite{Harlingen95}
To explain the mechanism of $d_{x^2-y^2}$-wave superconductivity, 
several mechanisms have been proposed.
Among others, the antiferromagnetic spin fluctuation theory
suggest a $d_{x^2-y^2}$-wave pairing mechanism\cite{Pines}
between d-orbital electrons at copper sites.
Another $d_{x^2-y^2}$-wave pairing mechanism is proposed in the
resonating valence bond (RVB) theory.\cite{ANDERSON}
In the RVB theory, electrons are assumed to be spin-charge separated.
In the slave-boson approach, \cite{LNW04}
spinons, which carry the spin degrees of
freedom, form a $d_{x^2-y^2}$-wave pairing state.
This pairing state is associated with the short-range
antiferromagnetic correlations.\cite{ABZH87}
Combined with Bose-Einstein condensation of holons, which carry the
electric charge, the pairing state of the spinons
leads to a $d_{x^2-y^2}$-wave superconducting state.

Both of these theories are based on a pairing state between
copper site electrons which produce antiferromagnetic correlations.
However, from the sign of the Hall coefficient\cite{Takagi89}
the charge carriers seem to be doped holes.
Optical conductivity measurements also support that the charge
carriers are doped holes because
the Drude weight in the optical conductivity\cite{Uchida91}
is proportional to the doped hole concentration.
If we assume that the charge carriers are doped holes, 
then it is natural to expect that high-temperature
superconductivity is based on a $d_{x^2-y^2}$-wave pairing mechanism 
between doped holes.
In this paper, we propose such a mechanism 
based on a picture for the doped holes.
The picture is that each hole induces a spin texture
that is characterized by a non-vanishing topological charge density
in the spin system through the suppression of the antiferromagnetic
correlations.

In Ref.\cite{TM_PRB05}, a half-skyrmion spin texture 
\cite{SD02,BASKARAN}
formation is discussed in the single-hole-doped cuprate 
within the non-linear sigma model(NLSM) description\cite{CHN} of the
Heisenberg antiferromagnet.
The formation of the half-skyrmion spin texture can be understood 
by using an analogy to a superfluid system:
Let us consider a two-dimensional boson system which exhibits
superfluidity in the ground state.
The Bose-Einstein condensate is described by a Gross-Pitaevskii
equation.\cite{FW}
If we suppress the condensate at some point $P$, then we see that
a vortex is formed around $P$ from the analysis of the
Gross-Pitaevskii equation.
Note that the field described by the Gross-Pitaevskii equation 
is a classical field.
However, 
the non-condensed component near the vortex core is described by
quantum bosons.
In case of a charged boson system, the condensate is locally
suppressed by a magnetic flux, or a vortex.

We can apply this result to the two-dimensional quantum
Heisenberg antiferromagnet. 
In the Schwinger boson mean field theory,\cite{AA88}
the N{\' e}el ordering is described by Bose-Einstein condensation of
the Schwinger bosons.\cite{SB_BEC}
If we assume that the doped hole forms a spin singlet with a
localized spin at a copper site, then the condensate is suppressed 
at that site.
Forming a Zhang-Rice singlet in the antiferromagnet plays a similar
role of introducing a magnetic flux in a charged boson condensate.
As in the case of a superfluid system, a vortex appears around the
site.
In terms of the spins, the vortex is a spin texture with the
non-vanishing topological charge density 
\cite{skyrmions}
that is given by
\begin{equation}
q({\bf r})=\frac{1}{4\pi} {\bf n}({\bf r}) \cdot \left[
\partial_x {\bf n} ({\bf r})\times 
\partial_y {\bf n} ({\bf r})\right],
\end{equation}
where the unit vector ${\bf n}({\bf r})$ describes the staggered
component of the spins.
Note that the spin at the core of the spin texture is a quantum spin
analogous to the normal component near a vortex core in a superfluid,
and so the spin can form a singlet state with the doped hole spin,
which is the Zhang-Rice singlet.\cite{ZHANG_RICE}
Based on the Schwinger boson mean field theory, we find that
the topological charge, which is obtained by integrating $q({\bf r})$
over a whole region, is quantized
to $\pm 1/2$.\cite{TM_PRB05,WENG_ETAL}
Because of this quantized value, the spin texture is called
half-skyrmion.

As argued in Ref.\cite{TM_PRB05}, the dispersion of the half-skyrmion
is in good agreement with the results of the angle-resolved
photoemission spectroscopy (ARPES) in the undoped
compounds.\cite{WELLS_ETAL}
The dispersion of the half-skyrmion excitation is the same as
that of the quasiparticles in the $\pi$-flux phase with a dynamically
induced mass.\cite{MARSTON,KIM_LEE}
In fact, the effective theory of the half-skyrmion is similar to that
of the $\pi$-flux phase:
the half-skyrmions are described by Dirac fermions with a U(1) gauge
field interaction.
Anomalously broad peaks observed in ARPES are associated with the
self-trapping of the half-skyrmions due to the coupling to the
longitudinal gauge field.\cite{TM_BROAD}

In Ref.\cite{TM_PRB05}, we assume that there is N{\' e}el ordering.
The presense of the N{\' e}el ordering is used for obtaining the quantized 
topological charge.
In the absence of the N{\' e}el ordering, the topological charge is no more 
quantized because the antiferromagnetic correlations 
decay exponentially with a length scale given by
the antiferromagnetic correlation length $\xi_{AF}$.
However, the system preserves
the local relationship between the hole density and the topological
charge density.
We consider spin textures with such topological charge density and 
hereafter we call such a spin texture staggered-spin-vortex.

The rest of the paper is organized as follows:
In Sec.\ref{sec_model}, we describe the model for 
the staggered-spin-vortex formation.
The pairing interaction is discussed in Sec.\ref{sec_pairing}.
We solve the gap equation in Sec.\ref{sec_gap} and show that the 
stable pairing state has $d_{x^2-y^2}$-wave symmetry.
Section \ref{sec_conclusion} is devoted to the conclusion.

\section{Model}
\label{sec_model}
We assume that the holes and the localized spins are independent
degrees of freedom.
Such a model is, for instance, the spin-fermion model, which is
derived from the d-p model by applying a canonical transformation.
In such a model, the spin system is described by the antiferromagnetic 
Heisenberg model, 
\begin{equation}
H_s=J\sum_{\langle i,j \rangle} {\bf S}_i \cdot {\bf S}_j.
\end{equation}
Here the summation is taken over the nearest neighbor sites and ${\bf
S}_i$ denote the spin moment at copper sites.
As an effective theory of the spin system, we consider the CP$^1$
model,\cite{RAJARAMAN} 
which is derived from the NLSM by representing ${\bf n}(x)$
in terms of the Schwinger bosons:
${\bf n} = \sum_{\sigma,\sigma'}\overline{z}_{\sigma} 
{\mbox{\boldmath ${\bf \sigma}$}}_{\sigma,\sigma'} z_{\sigma}$:
\begin{equation}
S_{CP^1} = \frac{2}{g}
\int d^3 x \sum_{\sigma=\uparrow,\downarrow}
\left[
\left| \left( \partial_{\mu}-i\alpha_{\mu} \right)
z_{\sigma} (x) \right|^2 
+ \frac{\Delta_{\rm sw}^2}{c_{\rm sw}^2}
|z_{\sigma} (x) |^2
\right],
\label{eq_CP1}
\end{equation}
where $\Delta_{\rm sw}$ is the antiferromagnetic spin wave excitation
gap and $c_{\rm sw}$ is the spin wave velocity.
The coupling $g$ is given by $g=c_{\rm sw}/\rho_s$ with
$\rho_s=Z_{\rho} J/4$ the spin stiffness constant.
$Z_{\rho}$ is the renormalization factor due to quantum phase
fluctuations.
In the spin disordered phase, $\Delta_{\rm sw}\neq 0$,
and there is no antiferromagnetic long-range order.
The antiferromagnetic correlations are characterized by
the correlation length 
$\xi_{AF} = c_{\rm sw}/\Delta_{\rm sw}$.
For the Schwinger bosons,
the presence of the antiferromagnetic correlations is associated with
the phase coherence among the Schwinger bosons.

For the kinetic energy of the holes, we assume 
\begin{equation}
H_h = \sum_{k\sigma} \epsilon_k c_{k\sigma}^{\dagger} c_{k\sigma},
\label{eq_ke}
\end{equation}
where $\epsilon_k = k^2/2m$ with $m$ the effective mass.
This is an approximate form in the continuum.

The coupling between the holes and the localized spins ${\bf S}_j$ 
has the form of the antiferromagnetic Kondo coupling:
\begin{equation}
H_K = J_K \sum_j \left(c_j^{\dagger} \sigma c_j \right)\cdot
{\bf S}_j.
\end{equation}
The parameter $J_K$ is on the order of $1\sim 3$eV.
Because of this strong Kondo coupling, holes suppress 
phase coherence of the Schwinger bosons
locally through the Zhang-Rice singlet formation.
The suppression of the phase coherence leads to a
staggered-spin-vortex:
\begin{equation}
\nabla \times 
{\mbox{\boldmath ${\bf \alpha}$}}
=
\pi \sum_{s=\pm} s \psi^{\dagger}_s ({\bf r}) \psi_s ({\bf r}),
\end{equation}
where 
$\nabla \times 
{\mbox{\boldmath ${\bf \alpha}$}}
= \partial_x \alpha_y - \partial_y \alpha_x$
and
$\psi^{\dagger}_{+(-)}$ is the creation operator of the
staggered-spin-vortex (anti-staggered-spin-vortex).
The coefficient is determined by considering the limit of
$\xi_{\rm AF} \rightarrow \infty$.
In this limit, the spin texture is the half-skyrmion.
The gauge flux $\nabla \times 
{\mbox{\boldmath ${\bf \alpha}$}}$ is 
associated with the spin chirality.\cite{LEE_NAGAOSA02,WWZ}
Note that $\langle \nabla \times
{\mbox{\boldmath ${\bf \alpha}$}}
\rangle =0$, 
because in the equilibrium there is an equal number of staggered spin
vortices and anti-staggered spin vortices, and so 
$\langle \sum_s s \psi^{\dagger}_s ({\bf r}) 
\psi_s ({\bf r}) \rangle=0$.

%%%%%%%%%%%%%%%%%%%%%%%%%%%%%%%%%%%%%%%%%%%%%%%%%%%%%%%
By integrating out the Schwinger bosons, we obtain the effective
action of the gauge field $\alpha_{\mu}$.
Since $\Delta_{\rm sw}\neq 0$, the action takes the Maxwellian form.
That is, the gauge field is massless.
However, in the spin disordered phase, the antiferromagnetic
correlation length $\xi_{AF}$ is finite.
Therefore, the gauge field propagator decays 
with a length scale of $\xi_{AF}$.
To include this feature, we perform a duality mapping.
We write $z_{\sigma}(x)$ as
\begin{equation}
z_{\sigma}(x)=\overline{\rho}^{1/2}_{\sigma} \exp(i\theta_0 +
i\theta_v).
\end{equation}
Here the phase $\theta_0$ is associated with the coherent motion of
the Schwinger bosons and the phase $\theta_v$ is associated with the
staggered-spin-vortices.
Note that this form is applicable for the description outside the
vortex cores.
After performing a Stratonovich-Hubbard transformation in
Eq.(\ref{eq_CP1}), we obtain the following Lagrangian density,
\begin{equation}
{\cal L} = \frac{g}{8\overline{\rho}} J_{\mu}^2 +iJ_{\mu} 
(\partial_{\mu} \theta_0 + \partial_{\mu} \theta_v - \alpha_{\mu}).
\end{equation}
We find $\partial_{\mu} J_{\mu}=0$ by integrating out $\theta_0$.
This constraint is satisfied by introducing a gauge field:
\begin{equation}
J_{\mu}=\frac{1}{2\pi} \epsilon_{\mu\nu\lambda} \partial_{\nu}
A_{\lambda}.
\end{equation}
We define the vortex current by
\begin{equation}
j_{\mu}^v = \frac{1}{2\pi} \epsilon_{\mu\nu\lambda} \partial_{\nu}
\partial_{\lambda} \theta_v.
\end{equation}
In terms of the staggered-spin-vortex 
and anti-staggered-spin-vortex fields, 
the vortex density is given by 
\begin{equation}
\rho^v ({\bf r}) = \sum_s s \psi^{\dagger}_s ({\bf r})
\psi_s ({\bf r}).
\end{equation}
From the equation of the continuity, $\partial_t \rho^v + \nabla \cdot 
{\bf j}^v =0$,
we find 
\begin{equation}
{\bf j}^v = 
\sum_s 
\frac{s}{2mi} \left[ \psi_s^{\dagger} \nabla \psi_s
- (\nabla \psi_s^{\dagger}) \psi_s 
\right],
\end{equation}
where we have used the fact that the kinetic energy of the holes is
described by Eq.(\ref{eq_ke}).
The coupling between the gauge field $A_{\mu}$ and the vortex current
$j_{\mu}^v$ is of the minimal coupling:
\begin{equation}
{\cal L}_{int} = -ij_{\mu}^v A_{\mu}.
\label{eq_coupling}
\end{equation}

Since $J_0=\nabla \times {\bf A}/2\pi$ and $J_0$ describes phase
fluctuations of the Schwinger boson density,
nonvanishing gauge flux $\nabla \times {\bf A}$ 
describes the suppression of 
the phase coherence.
Contrary to the gauge flux 
$\nabla \times 
{\mbox{\boldmath ${\bf \alpha}$}}$, 
the flux quantum for the Meissner phase of ${\bf A}$ is 
$2\pi$ because staggered-spin-vortex carry a unit gauge charge.
Therefore, the relation between the vortex density and the
gauge flux is
\begin{equation}
\sum_s \psi^{\dagger}_s ({\bf r})
\psi_s ({\bf r})
=-\frac{1}{2\pi} \nabla \times {\bf A}.
\label{eq_flux}
\end{equation}
The gauge field $A_{\mu}$ is equivalent to the gauge field in the
slave-boson mean field theory. \cite{LEE_NAGAOSA02}

\section{Pairing interaction}
\label{sec_pairing}
The pairing interaction is obtained from
Eqs.(\ref{eq_flux}) and (\ref{eq_coupling}) 
by eliminating the gauge field $A_{\mu}$.
We take the Coulomb gauge:
\begin{equation}
\nabla \cdot {\bf A} =0.
\end{equation}
In the momentum space,
\begin{equation}
A_{qx} = \frac{iq_y}{q^2} A_q,
\hspace{2em}
A_{qy} = -\frac{iq_x}{q^2} A_q.
\end{equation}
From Eq.(\ref{eq_flux}) we obtain
\begin{equation}
A_q = -2\pi \sum_{ks} \psi^{\dagger}_{ks} \psi_{k+q,s}.
\end{equation}
The staggered-spin-vortex current operator is
\begin{equation}
{\bf j}_{\bf q}^v = 
\sum_{{\bf k},s} 
\frac{s}{m} \left( {\bf k}+\frac{{\bf q}}{2} \right)
\psi^{\dagger}_{ks} \psi_{k+q,s}.
\end{equation}
Substituting these equations into Eq.(\ref{eq_coupling}), we obtain
\begin{equation}
H_{\rm int}=-\frac{2\pi i}{m\Omega}
\sum_{q ,s,s'}
\frac{{\bf q}\times {\bf k}'}{q^2}s' 
\psi^{\dagger}_{k+q,s} \psi_{k's'}^{\dagger}
\psi_{k'+q,s'}\psi_{ks},
\end{equation}
where $\Omega$ is the area of the system.
Since we are interested in the Cooper pairing, we set ${\bf k}+{\bf
k}'+{\bf q}=0$.
Thus, we obtain
\begin{equation}
H_{\rm int}=-\frac{2\pi i}{m\Omega}
\sum_{{\bf k}\neq{\bf k}',s,s'} 
\frac{{\bf k}\times {\bf k}'}{|{\bf k}-{\bf k}'|^2}
s'
\psi^{\dagger}_{k's} \psi_{-k',s'}^{\dagger}
\psi_{-k,s'}\psi_{ks}.
\end{equation}
By making replacement of $s\rightarrow s'$, 
$s'\rightarrow s$, ${\bf k}'\rightarrow -{\bf k}'$,
and ${\bf k}\rightarrow -{\bf k}$,
we obtain
\begin{equation}
H_{\rm int}=-\frac{2\pi i}{m\Omega}
\sum_{{\bf k}\neq{\bf k}',s,s'} 
\frac{{\bf k}\times {\bf k}'}{|{\bf k}-{\bf k}'|^2} s
\psi^{\dagger}_{k's} \psi_{-k',s'}^{\dagger}
\psi_{-k,s'}\psi_{ks}.
\end{equation}
Therefore, we may write
\begin{equation}
H_{\rm int}=- \frac{i\pi}{m\Omega}
\sum_{{\bf k}\neq{\bf k}',s,s'} 
\frac{{\bf k}\times {\bf k}'}{|{\bf k}-{\bf k}'|^2}(s+s')
\psi^{\dagger}_{k's} \psi_{-k',s'}^{\dagger}
\psi_{-k,s'}\psi_{ks}.
\label{eq_pairing}
\end{equation}
From this form, it is apparent that the interaction exists for either
staggered-spin-vortex pairs or anti-staggered-spin-vortex pairs.
The Hamiltonian is given by
\begin{equation}
H=\sum_s H^{(s)},
\label{eq_h}
\end{equation}
\begin{equation}
H^{(s)} = \sum_{\bf k} \epsilon_k \psi_{ks}^{\dagger}
\psi_{ks}
- \frac{2\pi i}{m\Omega} \sum_{{\bf k}\neq {\bf k}'}
\frac{{\bf k}\times {\bf k}'}{|{\bf k}-{\bf k}'|^2}
\psi^{\dagger}_{k's} \psi_{-k',s}^{\dagger}
\psi_{-k,s}\psi_{ks}.
\label{eq_hs}
\end{equation}

A similar pairing interaction was discussed in the composite fermion
system at half-filled Landau levels.\cite{GWW,TM_PRB00}.
In that system, the composite fermions are spinless fermion and there
is no other index associated with internal symmetry.
It was shown \cite{GWW2,TM_PRB00} that the pairing interaction leads
to $p_x \pm ip_y$-wave pairing state.
The sign is determined by the direction of the applied magnetic field
perpendicular to the system.
This pairing state is consistent with numerical
simulations.\cite{CF_NUM}

\begin{figure}[htbp]
\includegraphics[scale=0.3]{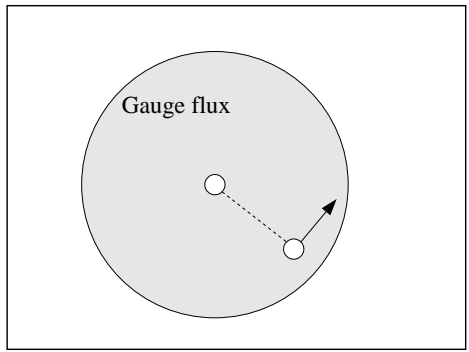}
\caption{Pairing interaction between the doped holes.
Shaded region represents the gauge flux $\nabla \times {\bf A}$
created by the hole sitting at the center.
If another hole passes this region with the Fermi velocity as shown in 
the figure, a magnetic Lorentz force like interaction is induced
between the two holes.
}
\label{fig_p}
\end{figure}

Before going into the analysis of the gap equation derived from the
Hamiltonian (\ref{eq_h}) and (\ref{eq_hs}), 
we show an intuitive view about the origin of the attractive
interaction.
According to Eq.~(\ref{eq_flux}), the gauge flux $\nabla \times {\bf
A}$
is produced by the doped holes.
Suppose a hole passes with the Fermi velocity the region of the gauge
flux created by another hole.
Then, as schematically shown in Fig.~\ref{fig_p}
the motion of the former hole is equivalent to a charged particle 
motion under a magnetic field.
Magnetic Lorentz force like interaction is induced between the two
holes.
Such an interaction leads to a chiral pairing state.
In this pairing mechanism, the gap is scaled by the Fermi energy.
Since holes carry either positive or negative gauge charge,
we expect there are two chiral pairing states with opposite
chiralities.
The stable pairing state in the bulk turns out to be
$d_{x^2-y^2}$-wave pairing state as we shall show in the next section.
%%%%%%%%%%%%%%%%%%%%%%%%%%%%%%%%%%%%%%%%%%%%%
% Figure fig_p.eps
%%%%%%%%%%%%%%%%%%%%%%%%%%%%%%%%%%%%%%%%%%%%%

\section{Gap equation}
\label{sec_gap}
Now we apply the BCS mean field theory to Eq.(\ref{eq_h}).
First we consider staggered-spin-vortex pairs described by $H^{(+)}$.
In the following, we set $\chi_k = \psi_{k+}$ and
$\chi_k^{\dagger}=\psi_{k+}^{\dagger}$,
to simplify notations.
We consider the grand canonical ensemble and
define the following mean field:
\begin{equation}
\Delta_k^{(+)} = -\frac{1}{\Omega}
\sum_{{\bf k}'(\neq {\bf k})}
V_{{\bf k}{\bf k}'} 
\langle \chi_{-{\bf k}'} \chi_{{\bf k}'}
\rangle,
\end{equation}
where
\begin{equation}
V_{{\bf k}{\bf k}'} = -\frac{4\pi i}{m}
\frac{{\bf k}\times {\bf k}'}{|{\bf k}-{\bf k}'|^2}.
\end{equation}
The mean field Hamiltonian is
\begin{equation}
H^{(+)}_{MF} = {\sum_{\bf k}}' \left[
\xi_k \chi_{k}^{\dagger} \chi_{k}
-\xi_k \chi_{-k} \chi_{-k}^{\dagger} 
+(\Delta_k^{(+)})^* \chi_{-k} \chi_{k}
+\Delta_k^{(+)} \chi^{\dagger}_{k} \chi^{\dagger}_{-k}
\right],
\end{equation}
where $\xi_k = \epsilon_k-\mu$ with $\mu$ the chemical potential and
the summation in the momentum space is taken over half of the
Brillouin zone.
The gap equation is given by
\begin{equation}
\Delta_{\bf k}^{(+)} = -\frac{1}{2\Omega}
\sum_{{\bf k}'(\neq {\bf k})}
V_{{\bf k}{\bf k}'}
\frac{\Delta_{{\bf k}'}^{(+)}}{E_{{\bf k}'}}
\tanh \frac{\beta E_{{\bf k}'}}{2},
\label{eq_gap}
\end{equation}
with $E_{\bf k}=\sqrt{\xi_k^2+|\Delta_{\bf k}^{(+)}|^2}$.

A similar gap equation is analyzed in Ref.\cite{GWW} in the context of 
the composite fermion pairing.
We apply the analysis presented in Ref.\cite{GWW}.
In order to solve the gap equation, we introduce the following
Ansatz\cite{GWW2}:
\begin{equation}
\Delta_{\bf k}^{(+)} = \Delta_k
\exp \left( -i\ell \theta_{\bf k} \right),
\end{equation}
where $\ell$ is an integer.
In Eq.~(\ref{eq_gap})
the integral with respect to $\theta_{{\bf k}'}$ is reduced to 
\begin{equation}
I_{\ell}(\eta)=\int_0^{2\pi} d\theta \frac{\sin \theta}{\cos \theta -\eta}
\exp (i\ell \theta),
\end{equation}
with $\eta=(k^2+k'^2)/(2kk')$.
The function $I_{\ell} (\eta)$ is exactly calculated by setting 
$z=\exp (i\theta)$
and applying a contour integral in the complex plane.
From Eq.~(\ref{eq_gap}), we see that if the interaction is attractive
(repulsive) then the sign of $I_{\ell}(\eta)$ is positive (negative).
For $\ell >0$ case, $I_{\ell}(\eta)>0$ while $\ell <0$ case,
$I_{\ell}(\eta)<0$, and $I_{\ell=0}(\eta)=0$.
Therefore, the gap equation has solutions only for $\ell>0$.
Thus, we obtain
\begin{equation}
\Delta_k = \frac{1}{2m}
\left[
\int_0^k dk' 
\frac{k'\Delta_{k'}}{E_{k'}}
\left( \frac{k'}{k} \right)^{\ell}
+ 
\int_k^{\infty} dk' \frac{k'\Delta_{k'}}{E_{k'}}
\left( \frac{k}{k'} \right)^{\ell} \right].
\end{equation}
From the asymptotic forms in $k\rightarrow \infty$ and 
$k\rightarrow 0$, we take the following approximate form:\cite{GWW}
\begin{equation}
\Delta_k = 
\left\{
\begin{array}{ccc}
\Delta \epsilon_F (k/k_F)^{\ell} & {\rm for} & k<k_F \\
\Delta \epsilon_F (k_F/k)^{\ell} & {\rm for} & k>k_F
\end{array}
\right.,
\end{equation}
where $\epsilon_F$ is the Fermi energy of the holes.
The gap $\Delta$ is obtained from the following equation:
\begin{equation}
\int_0^1 dx \frac{x^{2\ell+1}}{\sqrt{(x^2-1)^2+\Delta^2 x^{2\ell}}}
+ \int_1^{\infty} dx 
\frac{x^{1-2\ell}}{\sqrt{(x^2-1)^2+\Delta^2 x^{-2\ell}}}
=1.
\end{equation}
The largest gap $\Delta = 0.916$ is obtained for the case of
$p$-wave ($\ell=1$).
The second largest gap $\Delta =0.406$ is obtained for the case of
$d$-wave ($\ell=2$).
The third largest gap is $\Delta=0.264$ for $\ell=3$.

For anti-staggered-spin-vortex pairs, we may carry out the same
analysis.
Since the interaction term has opposite sign compared to the 
staggered-spin-vortex 
pair case, the gap function $\Delta_{\bf k}^{(-)}$ has the
following form:
\begin{equation}
\Delta_{\bf k}^{(-)} = \Delta_k
\exp \left( i\ell \theta_{\bf k} \right),
\end{equation}
with $\ell>0$.

From the above analyses, the linear combination of the gap functions
is
\begin{equation}
\Delta_{\bf k}=(\Delta_{\bf k}^{(+)}+\Delta_{\bf k}^{(-)})/2=
\Delta_k \cos \left( \ell \theta_{\bf k} \right).
\end{equation}
The relative phase between $\Delta_{\bf k}^{(+)}$
and $\Delta_{\bf k}^{(-)}$ is set to be zero.
(A non-zero phase can appear if there is a magnetic field.)
For the case of $\ell=1$, the symmetry of the Cooper pair is $p_x$.
Such a state can be stablized only at the boundary of the sample in
the square lattice.
Meanwhile, $\ell=2$ case, that is, $d_{x^2-y^2}$-wave pairing state is 
stable in the bulk.
%Although the possibility of $d_{xy}$-symmetry is not ruled out,
%such a state has a smaller gap than that of $d_{x^2-y^2}$ symmetry on
%the square lattice.
Thus, we may conclude that the above pairing interaction leads to 
$d_{x^2-y^2}$-wave superconductivity.

%spin
The above analysis is carried out for the staggered-spin-vortices.
In terms of those fields, the hole spin states are implicit.
For spin singlet pairing states, the spin states should be
\begin{equation}
\langle \chi_{-k}^{\downarrow} \chi_k^{\uparrow}
\rangle,
\end{equation}
and
\begin{equation}
\langle \chi_{-k}^{\uparrow} \chi_k^{\downarrow}
\rangle,
\end{equation}
where $\sigma$ in $\chi_k^{\sigma}$ denotes the hole spin state.
The relative phase, which is arbitrary in the above analysis, 
is $\pi$ for the spin-singlet pairing state:
\begin{equation}
\langle \chi_{-k}^{\downarrow} \chi_k^{\uparrow}
\rangle=
-\langle \chi_{-k}^{\uparrow} \chi_k^{\downarrow}
\rangle.
\end{equation}

Now we consider the effect of the staggered-spin vortex-vortex
interaction and the Coulomb interaction.
Those repulsive interactions reduce the gap value.
In the following we consider those effects separately.
%vortex-vortex interaction
In order to evaluate the vortex-vortex interaction effect, 
we consider a two-vortex solution as follows:
\begin{equation}
\theta_v = \tan^{-1} \frac{y}{x} + \tan^{-1} \frac{y}{x-d}.
\end{equation}
The interaction energy between the vortices is
\begin{equation}
V_{vv} (d) = \frac{J}{2}Z_{\rho} \overline{\rho}
\int d^2 {\bf r}
\frac{{\hat z}\times {\bf r}}{r^2}
\cdot
\frac{{\hat z}\times \left( x-d,y \right)}{(x-d)^2+y^2}.
\end{equation}
Taking into account the fact that the range of the interaction is
limited by the antiferromagnetic correlation length $\xi_{AF}$, we
obtain
\begin{equation}
V_{vv}(d) \simeq \pi J Z_{\rho} \overline{\rho} 
\ln \frac{\xi_{AF}}{d}.
\end{equation}
In the momentum space, 
\begin{equation}
V_{vv} (q) = \frac{2\pi\xi_{AF}^2}{q^2} J Z_{\rho} \overline{\rho},
\end{equation}
with the constraint, $q>2\pi/\xi_{AF}$.

\begin{figure}[htbp]
\includegraphics[scale=0.5]{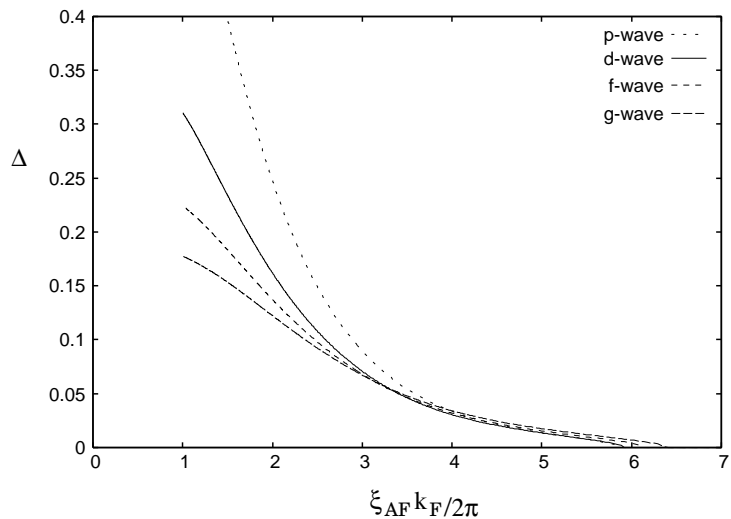}
\caption{The gap parameter $\Delta$ versus 
$\xi_{AF}k_F/(2\pi)$.}
\label{fig_vv}
\end{figure}

In the presence of the vortex-vortex interaction $V_{vv}(q)$, the gap
equation is given by
\begin{eqnarray}
1&=&\int_0^1 dx \frac{x^{2\ell+1}}{\sqrt{(x^2-1)^2+\Delta^2 x^{2\ell}}}
+ \int_1^{\infty} dx 
\frac{x^{1-2\ell}}{\sqrt{(x^2-1)^2+\Delta^2 x^{-2\ell}}}
\nonumber \\
& & - C_v 
\left[
\int_0^{1-\zeta} dx 
\frac{x^{2\ell}}{\sqrt{(x^2-1)^2+\Delta^2 x^{2\ell}}}
\frac{x}{1-x^2}
+ \int_{1+\zeta}^{\infty} dx 
\frac{x^{-2\ell}}{\sqrt{(x^2-1)^2+\Delta^2/x^{2\ell}}}
\frac{x}{x^2-1}
\right],
\end{eqnarray}
where $\zeta=2\pi/(k_F \xi_{AF})$ and the constant $C_v$ is
\begin{equation}
C_v = \frac{\pi J}{8\epsilon_F} Z_{\rho} \overline{\rho}\xi_{AF}^2.
\end{equation}
We take $C_v \simeq 0.5$ as a reasonable value at $x=0.10$ 
by setting 
$\overline{\rho} \xi_{AF}^2 \sim 1$,
$J/\epsilon_F \sim 2$. 
For $Z_{\rho}$, the value of $Z_{\rho}=0.72$ is used which is
estimated from quantum Monte Carlo simulations.\cite{QMC}
Figure \ref{fig_vv} shows $k_F \xi_{AF}$ dependence of the gap
parameters for each $\ell$.
%%%%%%%%%%%%%%%%%%%%%%%%%%%%%%%%%%%%%%%%%%%%%
% Figure fig_vv.eps
%%%%%%%%%%%%%%%%%%%%%%%%%%%%%%%%%%%%%%%%%%%%%
The effect of the vortex-vortex interaction is large
for strong antiferromagnetic correlations.
The gap values are somewhat reduced by the vortex-vortex interaction.
While the relevant parameter region would be
$\xi_{AF} k_F/(2\pi) = 1\sim 2$,
around $\xi_{AF} k_F/(2\pi)\sim 3$, 
the f-wave gap becomes larger than the d-wave gap.
Note that as an approximation, 
$C_v$ is fixed in the above calculation for simplicity.
For more precise calculations,
we require detail analysis of $\xi_{AF} 
k_F$ dependence of $\overline{\rho}$ and $Z_{\rho}$.
%vortex-anti-vortex
%If we consider the vortex-anti-vortex interaction, then there is
%another attractive interaction.
%However, the gap values associated with such an interaction are
%much smaller than the above gap values.

Another pairing mechanism based on merons, which presumably correspond
to the spin texture with non-zero topological density,
was discussed by Berciu and John.\cite{BJ99,BJ00}
From the analysis of an extended Hubbard model, where a
nearest-neighbor Coulomb repulsion is included, it was suggested that
meron-antimeron pairs lead to d-wave superconductivity.
In Ref.\cite{BJ99,BJ00}, the pairing interaction between merons and
antimerons comes from the vortex-antivortex interaction,
and the pairing interaction (\ref{eq_pairing}) is not considered.
Although the vortex-antivortex interaction is attractive interaction,
the analysis of the gap equation for the vortex-antivortex interaction 
within our Ansatz shows that there is no pairing state for 
$\xi_{AF} k_F/(2\pi) < 13$.
In this analysis, the coefficient $C_v$ is fixed.
More precise analysis requires $\xi_{AF}$ dependence of $C_v$.
However, $C_v$ is expected to be an increasing function with respect
to $\xi_{AF}$, and
the increase of $C_v$ leads to reduction of the gap values for large 
$\xi_{AF}$.

Now we consider the effect of the Coulomb interaction.
The Coulomb interaction between the holes is
\begin{equation}
V^C_q = \frac{2\pi e^2}{\epsilon q},
\end{equation}
with $\epsilon$ the dielectric constant.
The gap equation with the Coulomb interaction is
\begin{eqnarray}
1&=&\int_0^1 dx \frac{x^{2\ell+1}}{\sqrt{(x^2-1)^2+\Delta^2 x^{2\ell}}}
+ \int_1^{\infty} dx 
\frac{x^{1-2\ell}}{\sqrt{(x^2-1)^2+\Delta^2 x^{-2\ell}}}
\nonumber \\
& & - C_e
\left[
\int_0^{1} dx 
\frac{x^{2\ell+1/2}}{\sqrt{(x^2-1)^2+\Delta^2 x^{2\ell}}}
J_{\ell} \left(\frac{x^2+1}{2x} \right)
+ \int_1^{\infty} dx 
\frac{x^{1/2-2\ell}}{\sqrt{(x^2-1)^2+\Delta^2/x^{2\ell}}}
J_{\ell} \left(\frac{x^2+1}{2x} \right)
\right],
\end{eqnarray}
where
\begin{equation}
J_{\ell} (\eta)=\int_0^{2\pi} d\theta 
\frac{\cos (\ell \theta)}{\sqrt{\eta-\cos \theta}},
\end{equation}
\begin{equation}
C_e = \frac{e^2 k_F}{8\sqrt{2}\pi \epsilon}.
\end{equation}
%%%%%%%%%%%%%%%%%%%%%%%%%%%%%%%%%%%%%%%%%%%%%
% Figure fig_c.eps
%%%%%%%%%%%%%%%%%%%%%%%%%%%%%%%%%%%%%%%%%%%%%
\begin{figure}[htbp]
\includegraphics[scale=0.5]{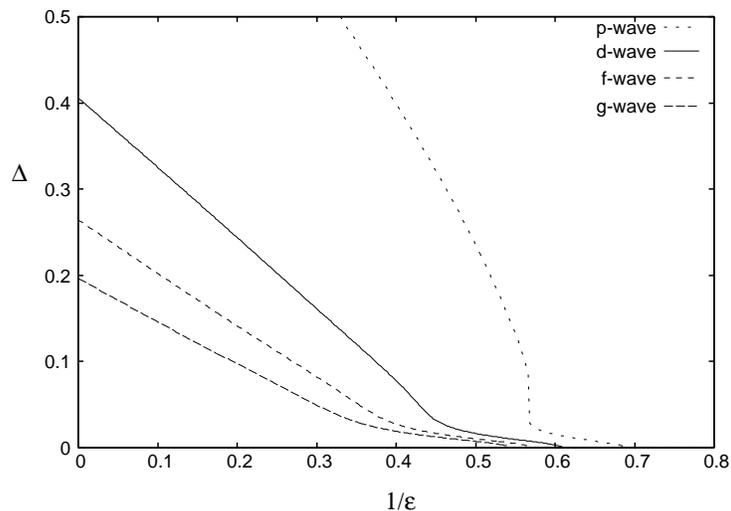}
\caption{The gap $\Delta$ versus $1/\epsilon$.}
\label{fig_c}
\end{figure}
Figure \ref{fig_c} shows $1/\epsilon$ dependence of the gap $\Delta$
for each $\ell$.

\section{Conclusion}
\label{sec_conclusion}
In this paper, we have proposed a mechanism of $d_{x^2-y^2}$-wave
superconductivity based on a spin texture with non-zero topological
charge density.
The spin texture formation is based on the Zhang-Rice singlet
formation in the background of the antiferromagnetic correlations.
In terms of a gauge field that describes antiferromagnetic spin
correlations, the spin texture is described by a gauge flux.
The interaction between the flux and the hole current induces
the magnetic Lorentz force like interaction between the holes.
Such an interaction leads to chiral pairing states with opposite
chiralities.
It turns out that the stable pairing state in the bulk is the
$d_{x^2-y^2}$-wave pairing state.
The stability of the pairing state is examined against the
vortex-vortex interaction and the Coulomb interaction.
%The appearance of the antiferromagnetic correlations is associated
%with the higher pseudogap temperature $T_0$.\cite{TIMUSK_STATT}
%While the formation of the Zhang-Rice singlet occurs at the Kondo
%temperature, $T_K(x)$.
%In the N{\' e}el ordering phase, $T_K(x)$ would be 
%$k_B T_K(x) \sim 1{\rm eV}$.
%However, in the spin disordered phase, $T_K(x)$ could be on the order
%of $100K$ because the magnitude of the localized moment at copper
%sites is significantly reduced.
%It would be interesting to consider the relationship between 
%$T_K(x)$ and the lower pseudogap temperature 
%$T^*$.\cite{TIMUSK_STATT}

In our pairing mechanism, the $p$-wave state is unstable in the bulk.
However, this state can be stabilized in the presence of anisotropy.
Since the $p$-wave gap is much larger than the d-wave gap, we expect
enhancement of the superconducting transition temperature if we can
induce that component.
If there is the $p$-wave component, then the parity is broken locally.
While the time-reversal symmetry is not broken unless one chiral
pairing state is suppressed.

\acknowledgments
I would like to thank Professor G. Baskaran for discussions and his
kind hospitality at Institute for Mathematical Sciences where part of
this work was done.
I also thank Dr. N. Nakai for discussions.
This work was supported by Grant-in-Aid for Young Scientists(B) 
(17740253) and the 21st Century COE "Center for Diversity and
Universality in Physics" from the Ministry of Education, Culture,
Sports, Science and Technology (MEXT) of Japan.

%\section{Duality mapping approach}
%\label{sec_duality}

\end{document}